\documentclass{cimento}
\usepackage[dvips]{graphicx}

 \title{An improved formula for the relativistic corrections to the kinematical Sunyaev-Zeldovich effect for clusters of galaxies}

 \shorttitle{Precision formula for the relativistic corrections}

 \author{Satoshi~Nozawa\from{ins:j}, Naoki~Itoh\from{ins:s}, Yasuhiko~Suda\from{ins:s} \atque Yoichi~Ohhata\from{ins:s}}

 \shortauthor{Nozawa et al.}

 \instlist{ \inst{ins:j} Josai Junior College, 1-1 Keyakidai, Sakado-shi, Saitama, 350-0290, Japan
 \inst{ins:s} Department of Physics, Sophia University, 7-1 Kioi-cho, Chiyoda-ku, Tokyo, 102-8554, Japan   }

\PACSes{ \PACSit{95.30.Cq}{Elementary particle processes}
         \PACSit{95.30.Gv}{Radiation mechanisms}
         \PACSit{98.65.Cw}{Galaxy clusters}
         \PACSit{98.80-k}{Cosmology}}

\begin{document}

\maketitle

\begin{abstract}

  We improve the calculation of Nozawa, Itoh, and Kohyama (1998) to provide a formula for relativistic corrections to the thermal and kinematical Sunyaev-Zeldovich effects that is accurate to fourth order in $\theta_{e} \equiv k_{B}T_{e}/m_{e}c^{2}$, $T_{e}$ and $m_{e}$ being the electron temperature and electron mass, respectively.  We also carry out a direct numerical integration of the Boltzmann collision term and confirm the excellent accuracy of the analytic formula.  This formula will be useful for the analysis of the observational data of the forthcoming experiments of the kinematical Sunyaev-Zeldovich effect for clusters of galaxies.

\end{abstract}

\section{Introduction}

  Compton scattering of the cosmic microwave background (CMB) radiation by hot intracluster gas --- the Sunyaev-Zeldovich effect \cite{zeld}--\cite{sz81} --- provides a useful method for studies of cosmology (see recent excellent reviews: \cite{birk99}, \cite{carl}).  Relativistic corrections to the Sunyaev-Zeldovich effect for cluster of galaxies have been extensively studied in recent years \cite{reph95}--\cite{noza05}.

  The present challenge to the observers of the Sunyaev-Zeldovich effect is the detection of the kinematical Sunyaev-Zeldovich effect, which is caused by the proper motion of galaxy clusters through CMB.  By observing the kinematical Sunyaev-Zeldovich effect one can deduce the velocity of the galaxy cluster along the line of sight.  Benson et al. \cite{bens03} have placed important upper limits to the velocities of the galaxy clusters along the line of sight with the method of the kinematical Sunyaev-Zeldovich effect.  Within the next 5 years CMB experiments such as PLANCK, SPT, ACT, QUIET, APEX and AMI will perform deep searches for galaxy clusters with sensitivity limits at the level of 1--10 mJy.  In the future CMB missions such as CMBPOL should reach sensitivities 20--100 times better than those of PLANCK by using currently existing technology \cite{chur02}.

  Under these current experimental circumstances it is worthwhile to obtain a precision formula for the relativistic corrections to the kinematical Sunyaev-Zeldovich effect for clusters of galaxies.  Nozawa, Itoh, \& Kohyama \cite{noza98} have calculated the relativistic corrections to the kinematical Sunyaev-Zeldovich effect for clusters of galaxies and have obtained a formula that is correct up to the order of $\beta \theta_{e}^{2}$ where $\theta_{e} \equiv k_{B}T_{e}/m_{e}c^{2}$, $T_{e}$ and $m_{e}$ being the electron temperature and electron mass, respectively, and $\beta=v/c$ is the magnitude of the peculiar velocity of the galaxy cluster with respect to CMB.  Sazonov \& Sunyaev \cite{sazo98} have calculated up to the $\beta \theta_{e}$ term with a different method; their result agreed with the $\beta \theta_{e}$ term obtained by Nozawa, Itoh, \& Kohyama \cite{noza98}.  Challinor \& Lasenby \cite{chal99} have carried out a further independent calculation and obtained the results that agreed perfectly with those of Nozawa, Itoh, \& Kohyama \cite{noza98} and Sazonov \& Sunyaev \cite{sazo98}.  More recently Shimon \& Rephaeli \cite{shim04} carried out an independent calculation of the relativistic corrections to the thermal and kinematical Sunyaev-Zeldovich effects.  However, their results for the kinematical Sunyaev-Zeldovich effect disagreed with those of Nozawa, Itoh, \& Kohyama \cite{noza98}, Sazonov \& Sunyaev \cite{sazo98}, and Challinor \& Lasenby \cite{chal99} even at the level of the $\beta \theta_{e}$ term.  In this paper we will carry out the calculation up to the order of $\beta \theta_{e}^{4}$ and obtain a precision formula for the relativistic corrections to the kinematical Sunyaev-Zeldovich effect for clusters of galaxies.  We will also carry out a direct numerical integration of the Boltzmann collision term and confirm the excellent accuracy of our analytic formula.  On the other hand, Shimon and Rephaeli's analytic formula shows considerable disagreement with the results of the direct numerical integration of the Boltzmann collision term.  Therefore we will conclude that Shimon and Rephaeli's analytic formula is in error.  We will discuss on this in sec. 3.

  The present paper is organized as follows.  In sec. 2 we briefly describe the method of the calculation and give the analytic solution.  In sec. 3 we show the numerical results of the calculation.  Concluding remarks will be given in sec. 4.  In Appendix we describe our method of the numerical integration.

\section{Lorentz boosted Kompaneets equation}

  In the present section we will follow Nozawa, Itoh, \& Kohyama \cite{noza98} and extend the Kompaneets equation to a system (the cluster of galaxies) moving with a peculiar velocity with respect to the CMB.  We will derive higher-order terms than Nozawa, Itoh, \& Kohyama \cite{noza98}.  We will formulate the kinetic equation for the photon distribution function using a relativistically covariant formalism.  As a reference system, we choose the system that is fixed to the CMB.  The $z$-axis is fixed to a line connecting the observer and the center of mass of the cluster of galaxies (CG). (We assume that the observer is fixed to the CMB frame.)  We fix the positive direction of the $z$-axis as the direction of the propagation of a photon from the cluster to the observer.  In this reference system, the center of mass of the cluster of galaxies is moving with a peculiar velocity $\vec{\beta} (\equiv \vec{v}/c$) with respect to the CMB.  For simplicity, we choose the direction of the velocity in the $x$-$z$ plane, i.e. $\vec{\beta} = (\beta_{x}, 0, \beta_{z})$.

  In the CMB frame, the time evolution of the photon distribution function $n(\omega)$ is written as follows:
\begin{eqnarray}
\frac{\partial n(\omega)}{\partial t} & = & -2 \int \frac{d^{3}p}{(2\pi)^{3}} d^{3}p^{\prime} d^{3}k^{\prime} \, W \,
\left\{ n(\omega)[1 + n(\omega^{\prime})] f(E) - n(\omega^{\prime})[1 + n(\omega)] f(E^{\prime}) \right\} \, ,  \\
W & = & \frac{(e^{2}/4\pi)^{2} \, \overline{X} \, \delta^{4}(p+k-p^{\prime}-k^{\prime})}{2 \omega \omega^{\prime} E E^{\prime}} \, ,  \\
\overline{X} & = & - \left( \frac{\kappa}{\kappa^{\prime}} + \frac{\kappa^{\prime}}{\kappa} \right) + 4 m^{4} \left( \frac{1}{\kappa} + \frac{1}{\kappa^{\prime}} \right)^{2} 
 - 4 m^{2} \left( \frac{1}{\kappa} + \frac{1}{\kappa^{\prime}} \right) \, ,  \\
\kappa & = & - 2 (p \cdot k) \, = \, - 2 \omega E \left( 1 - \frac{\mid \vec{p} \mid}{E} {\rm cos} \alpha \right) \, ,  \\
\kappa^{\prime} & = &  2 (p \cdot k^{\prime}) \, = \, 2 \omega^{\prime} E \left( 1 - \frac{\mid \vec{p} \mid}{E} {\rm cos} \alpha^{\prime} \right) \, .
\end{eqnarray}
In the above $W$ is the transition probability corresponding to the Compton scattering.  The four-momenta of the initial electron and photon are $p = (E, \vec{p})$ and $k = (\omega, 0, 0, k)$, respectively.  The four-momenta of the final electron and photon are $p^{\prime} = (E^{\prime}, \vec{p}^{\prime})$ and $k^{\prime} = (\omega^{\prime}, \vec{k}^{\prime})$, respectively.  The angles $\alpha$ and $\alpha^{\prime}$ are the angles between $\vec{p}$ and $\vec{k}$, and between $\vec{p}$ and $\vec{k}^{\prime}$, respectively.  Throughout this paper, we use the natural unit $\hbar = c = 1$ unit, unless otherwise stated explicitly.  Here we note that equation (1) assumes an isotropic photon distribution.  The kinematic effects lead to anisotropy but to a good approximation equation (1) may be applied all the way through the cluster since the levels of anisotropy induced by the kinematic effect are small.

  The electron distribution functions in the initial and final states are Fermi--like in the CG frame.  They are related to the electron distribution functions in the CMB frame as follows:
\begin{eqnarray}
f(E) & = &  f_{C}(E_{C})  \, , \\
f(E^{\prime}) & = &  f_{C}(E_{C}^{\prime})  \,  ,  \\
E_{C} & = & \gamma \, \left(E - \vec{\beta} \cdot  \vec{p} \right) \, ,   \\
E_{C}^{\prime} & = & \gamma \, \left(E^{\prime} - \vec{\beta} \cdot \vec{p}^{\prime} \right) \, ,  \\
\gamma & \equiv & \frac{1}{\sqrt{1 - \beta^{2}}}   \, ,
\end{eqnarray}
where the suffix $C$ denotes the CG frame.  We caution the reader that the electron distribution function is anisotropic in the CMB frame for $\beta \neq 0$.  By ignoring the degeneracy effects, we have the relativistic Maxwellian distribution for electrons with temperature $T_{e}$ as follows:
\begin{eqnarray}
f_{C}(E_{C}) & = & \left[ e^{\left\{(E_{C} - m)-(\mu_{C} - m) \right\}/k_{B}T_{e}} \, + \, 1 \right]^{-1}  \nonumber \\
& \approx & e^{-\left\{(E_{C}-m)-(\mu_{C} - m)\right\}/k_{B}T_{e}} \, ,
\end{eqnarray}
where $(\mu_{C} - m)$ is the non-relativistic chemical potential of the electron measured in the CG frame.  We now introduce the quantities
\begin{eqnarray}
x &  \equiv &  \frac{\omega}{k_{B}T_{e}}  \, ,  \\
\Delta x &  \equiv &  \frac{\omega^{\prime} - \omega}{k_{B}T_{e}}  \, .
\end{eqnarray}
Substituting equations (6) -- (13) into equation (1), we obtain
\begin{eqnarray}
\frac{\partial n(\omega)}{\partial t} = -2 \int \frac{d^{3}p}{(2\pi)^{3}} d^{3}p^{\prime} d^{3}k^{\prime} \, W \, f_{C}(E_{C}) \,
\left[ \, \left\{ \, 1 + n(\omega^{\prime}) \, \right\} n(\omega) \,
  \right.  \hspace{3.0cm}  \nonumber  \\
 \left. \, - \,  \left\{ \, 1 + n(\omega) \, \right\} n(\omega^{\prime}) \, {\rm e}^{ \Delta x \gamma (1 - \vec{\beta} \cdot \hat{k}^{\prime} ) } \, {\rm e}^{ x \gamma \vec{\beta} \cdot ( \hat{k} - \hat{k}^{\prime} ) } \right] \, ,
\end{eqnarray}
where $\hat{k}$ and $\hat{k}^{ \prime}$ are the unit vectors in the directions of $\vec{k}$ and $\vec{k}^{ \prime}$, respectively.  Equation (14) is our basic equation.

   We now expand equation (14) in powers of $\Delta x$ by assuming $\Delta x  \, \ll 1$.  We obtain the Fokker-Planck expansion
\begin{eqnarray}
\frac{ \partial n(\omega)}{ \partial t} & = & 
 2 \left[ \frac{ \partial n}{ \partial x} \, I_{1,0} + n(1+n) \, I_{1,1} \right]
  \nonumber  \\
& + & 2 \left[ \frac{ \partial^{2} n}{ \partial x^{2}} \, I_{2,0}
+ 2(1+n) \frac{ \partial n}{ \partial x} \, I_{2,1} + n(1+n)  \, I_{2,2} \right]
  \nonumber  \\
& + & 2 \left[\frac{ \partial^{3} n}{ \partial x^{3}} \, I_{3,0}
+ 3(1+n) \frac{ \partial^{2} n}{ \partial x^{2}} \, I_{3,1}
+ 3(1+n) \frac{ \partial n}{ \partial x} \, I_{3,2} + n(1+n) \, I_{3,3} \right]
  \nonumber \\
& + & \cdot \cdot \cdot  \,  \nonumber \\
& + & 2 \, n \, \left[ (1 + n) J_{0} + \frac{ \partial n}{ \partial x } \,  J_{1} + \frac{ \partial^{2} n}{ \partial x^{2}} \, J_{2} + \frac{ \partial^{3} n}{ \partial x^{3}} \, J_{3} + \cdot \cdot \cdot  \, \, \,  \right] \, \, ,
\end{eqnarray}
where
\begin{eqnarray}
I_{k, \ell} & \equiv & \frac{1}{k !} \int \frac{d^{3}p}{(2\pi)^{3}} d^{3}p^{\prime} d^{3}k^{\prime} \, W \, f_{C}(E_{C}) \, (\Delta x)^{k}  \,  {\rm e}^{ x \gamma \vec{\beta} \cdot ( \hat{k} - \hat{k}^{\prime} ) }  
\gamma^{ \ell} \left( 1 - \vec{\beta} \cdot \hat{k}^{\prime} 
\right)^{ \ell}  \, , \\
  \nonumber \\
J_{k} & \equiv & \frac{-1}{k !} \int \frac{d^{3}p}{(2\pi)^{3}} d^{3}p^{\prime} d^{3}k^{\prime} \, W \, f_{C}(E_{C}) \, (\Delta x)^{k}  \left( \, 1 \, - \, {\rm e}^{ x \gamma \vec{\beta} \cdot ( \hat{k} - \hat{k}^{\prime} ) }  \right)  \, .
\end{eqnarray}
Analytic integration of equations (16) and (17) can be done with the power series expansion approximation of the integrand in terms of the electron momentum $p$.  In Itoh et al. \cite{itoh98}, the systematic analysis has been done in order to examine the accuracy of the power series expansion approximation.  It has been found that the power series expansion approximation is sufficiently accurate for $k_{B}T_{e} \leq 15$keV by taking into account $O(\theta_{e}^{5})$ corrections.

In addition to $\theta_{e}$, there is another parameter $\vec{\beta}$ in equations (16) and (17).  For most of the cluster of galaxies, $\beta \ll 1$ is realized.  For example, $\beta \approx $ 1/300 for a typical value of the peculiar velocity $v$=1,000 kms$^{-1}$.  Therefore it should be sufficient to expand equations (16) and (17) in powers of $\beta$ and to retain up to  $O(\beta^{2})$ contributions.  We assume the initial photon distribution of the CMB to be Planckian with a temperature $T_{0}$:
\begin{equation}
n_{0} (X) \, = \, \frac{1}{e^{X} - 1} \, , 
\end{equation}
where
\begin{equation}
X \, \equiv \, \frac{\hbar \omega}{k_{B} T_{0}}  \, .
\end{equation}

Substituting the results of equations (16) and (17) into equation (15) and assuming $T_{0}/T_{e} \ll 1$, one obtains the following expression for the fractional distortion of the photon spectrum:
\begin{eqnarray}
\frac{\Delta n(X)}{n_{0}(X)} & = & \frac{\tau \, X e^{X}}{e^{X}-1} \, \theta_{e} \, \left[  \, \,
Y_{0} \, + \, \theta_{e} Y_{1} \, + \, \theta_{e}^{2} Y_{2} \, + \,  \theta_{e}^{3} Y_{3} \, + \, \theta_{e}^{4} Y_{4} \,  \right]  \,   \nonumber  \\
  & + & \frac{\tau \, X e^{X}}{e^{X}-1}  \, \beta^{2} \, \left[ \, \, B_{0} \, + \, \theta_{e} B_{1} \, + \, \theta_{e}^{2} B_{2} \, + \, \theta_{e}^{3} B_{3} \, \right]    \, \nonumber \\ 
  & + & \frac{\tau \, X e^{X}}{e^{X}-1}  \, \beta \, P_{1}({\rm cos} \theta_{\gamma}) \, \left[ \, \, C_{0} \, + \, \theta_{e} C_{1} \, + \, \theta_{e}^{2} C_{2} \, + \, \theta_{e}^{3} C_{3} \, + \, \theta_{e}^{4} C_{4} \,  \right]  \, \nonumber \\
& + & \frac{\tau \, X e^{X}}{e^{X}-1} \beta^{2}  P_{2} ({\rm cos} \theta_{\gamma}) \, \left[ \, D_{0} \, + \, \theta_{e} D_{1} \, + \, \theta_{e}^{2} D_{2} \, + \, \theta_{e}^{3} D_{3} \, \right] \, , 
\end{eqnarray}
\begin{eqnarray}
\tau & \equiv & \sigma_{T} \int d \ell N_{e}  \, , \\
\theta_{e} & \equiv & \frac{k_{B} T_{e}}{m_{e}c^{2}} \, , \\
{\rm cos} \theta_{\gamma} & = & \frac{\beta_{z}}{\beta} \, , \\
P_{1} ({\rm cos} \theta_{\gamma}) & = & {\rm cos} \theta_{\gamma} \, , \\
P_{2} ({\rm cos} \theta_{\gamma}) & = & \frac{1}{2} \, \left( 3 {\rm cos}^{2} \theta_{\gamma} - 1 \right) \, ,
\end{eqnarray}
where $\theta_{\gamma}$ is the angle between the directions of the peculiar velocity of the cluster $\vec{\beta}= \vec{v}/c$ and the photon momentum $\vec{k}$, which is chosen as the positive $z$-direction.  The reader should remark that this sign convention for the positive $z$-direction is opposite to the ordinary one.  Thus a cluster moving away from the observer has ${\rm cos} \theta_{\gamma} < 0$.  However, because of the positive sign in front of the $P_{1}({\rm cos} \theta_{\gamma})$ term, one obtains $\Delta n(X) < 0$ in this case, as one should.  The coefficients are defined as follows:
\begin{eqnarray}
Y_{0} & = & - 4 \, + \tilde{X}  \,  , \\
Y_{1} & = & - 10 + \frac{47}{2} \tilde{X} - \frac{42}{5} \tilde{X}^{2} + \frac{7}{10} \tilde{X}^{3}  \, + \, \tilde{S}^{2} \left( - \frac{21}{5} + \frac{7}{5} \tilde{X} \right) \,  ,  \\
Y_{2} & = & - \frac{15}{2} + \frac{1023}{8} \tilde{X} - \frac{868}{5} \tilde{X}^{2} + \frac{329}{5} \tilde{X}^{3} - \frac{44}{5} \tilde{X}^{4} + \frac{11}{30} \tilde{X}^{5}  \nonumber \\ 
& & + \tilde{S}^{2} \left( - \frac{434}{5} + \frac{658}{5} \tilde{X}  - \frac{242}{5}  \tilde{X}^{2} + \frac{143}{30} \tilde{X}^{3} \right) 
 +  \tilde{S}^{4} \left( - \frac{44}{5} + \frac{187}{60} \tilde{X} \right) \, ,   \\
Y_{3} & = & \frac{15}{2} + \frac{2505}{8} \tilde{X} - \frac{7098}{5} \tilde{X}^{2} + \frac{14253}{10} \tilde{X}^{3} - \frac{18594}{35} \tilde{X}^{4}   \nonumber  \\
& + &  \frac{12059}{140} \tilde{X}^{5} - \frac{128}{21} \tilde{X}^{6} + \frac{16}{105} \tilde{X}^{7} \nonumber \\ 
& + & \tilde{S}^{2} \left( - \frac{7098}{10} + \frac{14253}{5} \tilde{X} - \frac{102267}{35}  \tilde{X}^{2} + \frac{156767}{140} \tilde{X}^{3} - \frac{1216}{7}  \tilde{X}^{4} + \frac{64}{7} \tilde{X}^{5} \right)  \nonumber  \\
& + &  \tilde{S}^{4} \left( - \frac{18594}{35} + \frac{205003}{280} \tilde{X} - \frac{1920}{7}  \tilde{X}^{2} + \frac{1024}{35} \tilde{X}^{3} \right) \nonumber  \\
& + &  \tilde{S}^{6} \left( - \frac{544}{21} + \frac{992}{105} \tilde{X} \right) \, , \\
Y_{4} & = & - \frac{135}{32} + \frac{30375}{128} \tilde{X} - \frac{62391}{10} \tilde{X}^{2} + \frac{614727}{40} \tilde{X}^{3} - \frac{124389}{10} \tilde{X}^{4}   \nonumber  \\
& + &  \frac{355703}{80} \tilde{X}^{5} - \frac{16568}{21} \tilde{X}^{6} + \frac{7516}{105} \tilde{X}^{7} - \frac{22}{7} \tilde{X}^{8} + \frac{11}{210} \tilde{X}^{9} \nonumber \\ 
& + & \tilde{S}^{2} \left( - \frac{62391}{20} + \frac{614727}{20} \tilde{X} - \frac{1368279}{20} \tilde{X}^{2} + \frac{4624139}{80} \tilde{X}^{3} - \frac{157396}{7}  \tilde{X}^{4}  \right. \nonumber  \\
&  & \, \, \, \, \, + \, \left. \frac{30064}{7} \tilde{X}^{5} - \frac{2717}{7} \tilde{X}^{6} + \frac{2761}{210} \tilde{X}^{7}   \right)  \nonumber  \\
& + &  \tilde{S}^{4} \left( - \frac{124389}{10} + \frac{6046951}{160} \tilde{X} - \frac{248520}{7} \tilde{X}^{2} + \frac{481024}{35} \tilde{X}^{3} - \frac{15972}{7} \tilde{X}^{4}  \right. \nonumber  \\
&  &  \, \, \, \, + \, \left. \frac{18689}{140} \tilde{X}^{5}  \right) \nonumber  \\
& + &  \tilde{S}^{6} \left( - \frac{70414}{21} + \frac{465992}{105} \tilde{X} - \frac{11792}{7} \tilde{X}^{2} + \frac{19778}{105} \tilde{X}^{3} \right) \nonumber  \\
& + &  \tilde{S}^{8} \left( - \frac{682}{7} + \frac{7601}{210} \tilde{X} \right) \, , \\
B_{0} & = & \frac{1}{3} Y_{0} \, , \\
B_{1} & = & \frac{5}{6}Y_{0} + \frac{2}{3} Y_{1} \, , \\
B_{2} & = & \frac{5}{8} Y_{0} + \frac{3}{2} Y_{1} + Y_{2} \, , \\
B_{3} & = & -\frac{5}{8} Y_{0} + \frac{5}{4} Y_{1} + \frac{5}{2} Y_{2} + \frac{4}{3} Y_{3}  \, , \\
C_{0} & = & 1 \, , \\
C_{1} & = & 10 - \frac{47}{5} \tilde{X} + \frac{7}{5} \tilde{X}^{2} + \frac{7}{10} \tilde{S}^{2}  \,  ,  \\
C_{2} & = & 25 - \frac{1117}{10} \tilde{X} + \frac{847}{10} \tilde{X}^{2} - \frac{183}{10} \tilde{X}^{3} + \frac{11}{10} \tilde{X}^{4}    \nonumber \\ 
& & + \tilde{S}^{2} \left( \frac{847}{20} - \frac{183}{5} \tilde{X}  + \frac{121}{20}  \tilde{X}^{2} \right)  +  \frac{11}{10} \tilde{S}^{4}  \, ,   \\
C_{3} & = &  \frac{75}{4} - \frac{21873}{40} \tilde{X} + \frac{49161}{40} \tilde{X}^{2} - \frac{27519}{35} \tilde{X}^{3} + \frac{6684}{35} \tilde{X}^{4} \, \nonumber \\
&  & - \frac{3917}{210} \tilde{X}^{5} + \frac{64}{105} \tilde{X}^{6}  \, \nonumber \\
& + &  \tilde{S}^{2} \left( \frac{49161}{80} - \frac{55038}{35} \tilde{X} + \frac{36762}{35} \tilde{X}^{2} - \frac{50921}{210} \tilde{X}^{3} + \frac{608}{35} \tilde{X}^{4} \right) \, \nonumber \\
& + &  \tilde{S}^{4} \left( \frac{6684}{35} - \frac{66589}{420} \tilde{X} + \frac{192}{7} \tilde{X}^{2} \right) \, \nonumber \\
& + &  \frac{272}{105} \tilde{S}^{^6}  \, ,  \\
C_{4} & = & - \frac{75}{4} - \frac{10443}{8} \tilde{X} + \frac{359079}{40} \tilde{X}^{2} - \frac{938811}{70} \tilde{X}^{3} + \frac{261714}{35} \tilde{X}^{4} \, \nonumber \\
& & - \frac{263259}{140} \tilde{X}^{5} + \frac{4772}{21} \tilde{X}^{6} - 
  \frac{1336}{105} \tilde{X}^{7} + \frac{11}{42} \tilde{X}^{8} \, \nonumber \\
& + & \tilde{S}^{2} \left( \frac{359079}{80} - \frac{938811}{35} \tilde{X} + \frac{1439427}{35} \tilde{X}^{2} - \frac{3422367}{140} \tilde{X}^{3} + \frac{45334}{7} \tilde{X}^{4} \right. \, \nonumber \\
& & \left. - \frac{5344}{7} \tilde{X}^{5} + \frac{2717}{84} \tilde{X}^{6} \right)  \, \nonumber \\
& + &  \tilde{S}^{4} \left( \frac{261714}{35} - \frac{4475403}{280} \tilde{X} + \frac{71580}{7} \tilde{X}^{2} - \frac{85504}{35} \tilde{X}^{3} + \frac{1331}{7} \tilde{X}^{4} \right)   \, \nonumber \\
& + & \tilde{S}^{6} \left( \frac{20281}{21} - \frac{82832}{105} \tilde{X} + \frac{2948}{21} \tilde{X}^{2} \right)  \, \nonumber  \\
& + &  \frac{341}{42} \tilde{S}^{8}  \, \\
D_{0} & = & - \frac{2}{3} + \frac{11}{30} \tilde{X} \, , \\
D_{1} & = & - 4 + 12 \tilde{X} - 6 \tilde{X}^{2} + \frac{19}{30} \tilde{X}^{3}  \, + \, \tilde{S}^{2} \left( - 3 + \frac{19}{15} \tilde{X} \right) \,  ,  \\
D_{2} & = & -10 + \frac{542}{5} \tilde{X} - \frac{843}{5} \tilde{X}^{2} + \frac{10603}{140} \tilde{X}^{3} - \frac{409}{35} \tilde{X}^{4} + \frac{23}{42} \tilde{X}^{5}  \, \nonumber  \\
& + &  \tilde{S}^{2} \left( - \frac{843}{10} + \frac{10603}{70} \tilde{X} - 
     \frac{4499}{70} \tilde{X}^{2} + \frac{299}{42} \tilde{X}^{3} \right) \, \nonumber \\
& + & \tilde{S}^{4} \left( - \frac{409}{35} + \frac{391}{84} \tilde{X} \right) \, ,  \\
D_{3} & = & - \frac{15}{2} + \frac{4929}{10} \tilde{X} - \frac{39777}{20} \tilde{X}^{2} + \frac{1199897}{560} \tilde{X}^{3} - \frac{4392}{5} \tilde{X}^{4} \, \nonumber  \\
& & + \frac{16364}{105} \tilde{X}^{5} - \frac{3764}{315} \tilde{X}^{6} + \frac{101}{315} \tilde{X}^{7}  \, \nonumber  \\
& + &  \tilde{S}^{2} \left( - \frac{39777}{40} + \frac{1199897}{280} \tilde{X} - \frac{24156}{5} \tilde{X}^{2} + \frac{212732}{105} \tilde{X}^{3} - \frac{35758}{105} \tilde{X}^{4}  \right.  \, \nonumber  \\
& & \left.  + \frac{404}{21} \tilde{X}^{5} \right)  \, \nonumber  \\
& + &  \tilde{S}^{4} \left( - \frac{4392}{5} + \frac{139094}{105} \tilde{X} - 
     \frac{3764}{7} \tilde{X}^{2} + \frac{6464}{105} \tilde{X}^{3} \right)  \, \nonumber  \\
& + &  \tilde{S}^{6} \left( - \frac{15997}{315} + \frac{6262}{315} \tilde{X} \right)  \, ,  \\
\tilde{X} & \equiv &  X \, {\rm coth} \left( \frac{X}{2} \right)  \, , \\
\tilde{S} & \equiv & \frac{X}{ \displaystyle{ {\rm sinh} \left( \frac{X}{2} \right)} }   \, .
\end{eqnarray}
In the above, $\omega$ is the photon frequency, $T_{0}$ is the CMB temperature, $\sigma_{T}$ is the Thomson cross section, $N_{e}$ is the electron number density in the cluster frame, and the integral in equation (21) is over the photon path length in the cluster.

  Concerning the photon number conservation of equation (20), the following remarks should be noted: As for the thermal Sunyaev-Zeldovich effect terms, the photon number conservation is satisfied as discussed in Itoh, Kohyama \& Nozawa \cite{itoh98}.  Namely, $Y_{0}$, $Y_{1}$, $Y_{2}$, $Y_{3}$, and $Y_{4}$ terms separately vanish by the integration $\int dX X^{2}$.  Therefore the first and second lines vanish by the integration.  On the other hand, the third and fourth lines are proportional to $P_{1}({\rm cos} \theta_{\gamma})$ and $P_{2}({\rm cos} \theta_{\gamma})$, respectively.  Therefore the third and fourth lines vanish by the integration over the solid angle $\int d \Omega_{\gamma}$.

Finally, we define the distortion of the spectral intensity as follows:
\begin{eqnarray}
\Delta I & \equiv & \frac{X^{3}}{e^{X}-1} \frac{\Delta n(X)}{n_{0}(X)} \, .
\end{eqnarray}
Here we note that Itoh, Kohyama, \& Nozawa \cite{itoh98} have carried out the exact numerical integration of the Boltzmann equation for the thermal Sunyaev-Zeldovich effect and have compared the results with those obtained by the generalized Kompaneets equation, which correspond to the first line in equation (20).  They have thereby confirmed that the analytic expansion result including up to the $\theta_{e}^{4}Y_{4}$ term agrees with the exact result within the accuracy of about 1\% for all the values of $X$ for the electron temperature $k_{B}T_{e} \leq 15$keV ($\theta_{e} \leq 0.03$).

\section{Numerical results}

  We now study the higher-order terms of the relativistic corrections to the kinematical Sunyaev-Zeldovich effect.  In fig.\ 1 we have plotted the distortion of the spectral intensity $\Delta I/\tau$ as a function of $X$ for $k_{B}T_{e}$=10keV and $\beta=\beta_{z}=1/300$.  The thermal and kinematical Sunyaev-Zeldovich effects have been plotted.  For the kinematical Sunyaev-Zeldovich effect, the absolute value of the effect is shown.  For a receding cluster one has $\Delta I < 0$, whereas for an approaching cluster one has $\Delta I > 0$.  We have shown the decomposition of the absolute value of the kinematical Sunyaev-Zeldovich contribution in fig.\ 2.  The dotted curve is the leading order contribution originally derived by Sunyaev \& Zeldovich \cite{sz80a}, \cite{sz80b}.  The dashed curve is the result of Nozawa, Itoh, \& Kohyama \cite{noza98}, which includes up to $O(\beta \theta_{e}^{2})$ terms.  The dot-dashed curve is the present result that includes up to $O(\beta \theta_{e}^{3})$ terms.  The solid curve is the present result that includes up to $O(\beta \theta_{e}^{4})$ terms.  The triple-dot-dashed curve is the contribution of the $O(\beta^{2})$ terms, where a factor 10 was multiplied in order to make the curve visible in this figure.  It is clear from fig.\ 2 that the convergence of the power series expansion is extremely good at this temperature for all $X$ values.  It is also clear from fig.\ 2 that the $O(\beta^{2})$ correction is very small and it can be safely neglected.

\begin{figure}
\begin{center}
\includegraphics[angle=-90,width=0.8\textwidth]{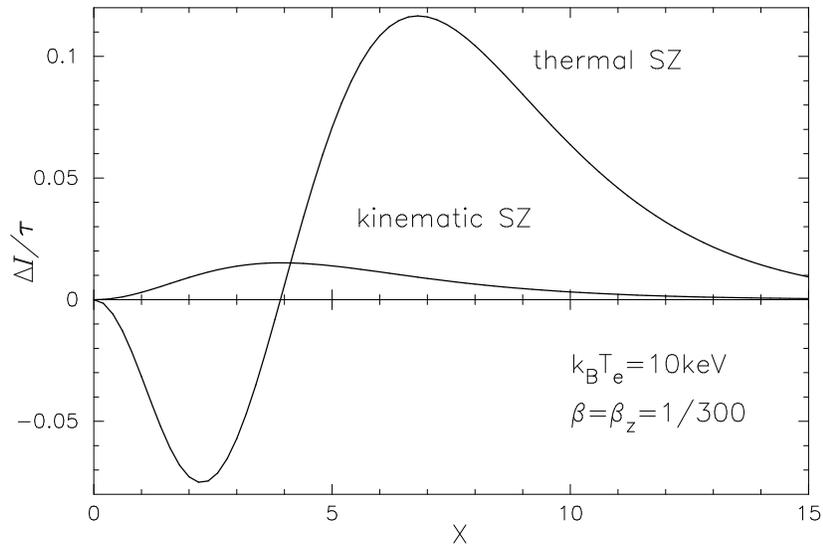}
\end{center}
\caption{Spectral intensity distortion $\Delta I/ \tau$ as a function of $X$ for $k_{B}T_{e}$ = 10keV, $\beta=\beta_{z}$=1/300.  The thermal and kinematical Sunyavev-Zeldovich effects have been plotted.  For the kinematical Sunyaev-Zeldovich effect, the absolute value of the effect is shown.}
\end{figure}

\begin{figure}
\begin{center}
\includegraphics[angle=-90,width=0.8\textwidth]{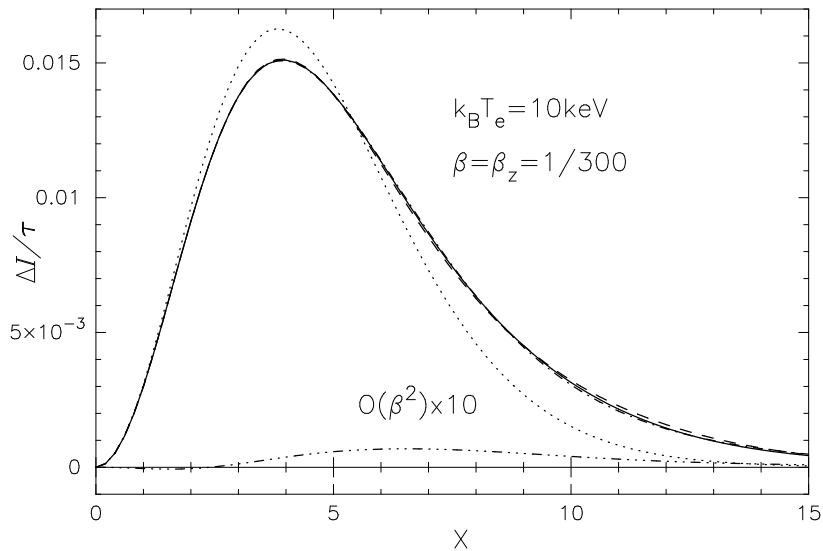}
\end{center}
\caption{The decomposition of the absolute value of the kinematical Sunyaev-Zeldovich contribution for $k_{B}T_{e}$ = 10keV, $\beta=\beta_{z}$=1/300.  The dotted curve is the leading order contribution originally derived by Sunyaev \& Zeldovich \cite{sz80a}, \cite{sz80b}.  The dashed curve is the result of Nozawa, Itoh, \& Kohyama \cite{noza98}, which includes up to $O(\beta \theta_{e}^{2})$ terms.  The dot-dashed curve is the present result that includes up to $O(\beta \theta_{e}^{3})$ terms.  The solid curve is the present result that includes up to $O(\beta \theta_{e}^{4})$ terms.  The triple-dot-dashed curve is the contribution of the $O(\beta^{2})$ terms, where a factor 10 was multiplied in order to make the curve visible in this figure.}
\end{figure}

It is well known that the kinematical Sunyaev-Zeldovich effect becomes extremely important in the crossover frequency region, where the thermal Sunyaev-Zeldovich effect vanishes.  Therefore the accurate position of the crossover frequency $X_{0}$ is extremely important.  In Itoh, Kohyama, \& Nozawa \cite{itoh98}, the following fitting formula for $X_{0}$ has been obtained.
\begin{equation}
X_{0} \, = \, 3.830 \, \left( \, 1 + \, 1.1674 \theta_{e} \, - \, 0.8533 \theta_{e}^{2} \, \right)  \, .
\end{equation}
The errors of this fitting function were less than $1 \times 10^{-3}$ for $0 \leq k_{B}T_{e} \leq 50{\rm keV}$.

Substituting equation (47) into the third line of equation (20), we have calculated the kinematical Sunyaev-Zeldovich effect at $X$ = $X_{0}$ as a function of $k_{B} T_{e}$ for $\beta=\beta_{z}$=1/300.  The result has been plotted in fig.\ 3.  The dotted curve is the result of Nozawa, Itoh, \& Kohyama \cite{noza98}, which includes up to $O(\beta \theta_{e}^{2})$ terms.  The dashed curve is the present result that includes up to $O(\beta \theta_{e}^{3})$ terms.  The dot-dashed curve is the present result that includes up to $O(\beta \theta_{e}^{4})$ terms.  The solid curve is the present result of the direct numerical integration.  It is clear from fig.\ 3 that the convergence of the power series expansion is extremely good for the temperature $k_{B}T_{e} \leq 20$keV.

\begin{figure}
\begin{center}
\includegraphics[angle=-90,width=0.8\textwidth]{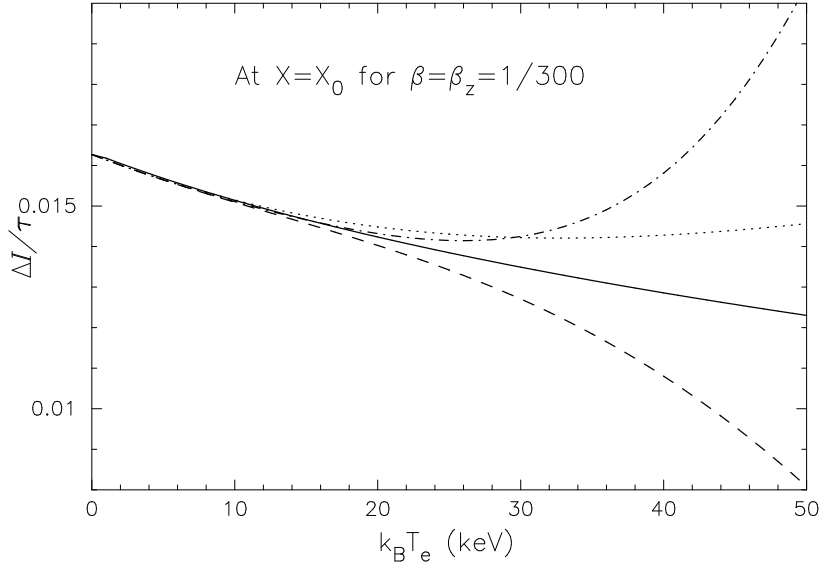}
\end{center}
\caption{The absolute value of the kinematical Sunyaev-Zeldovich effect at $X$ = $X_{0}$ as a function of $k_{B} T_{e}$ for $\beta=\beta_{z}$=1/300.  The dotted curve is the result of Nozawa, Itoh, \& Kohyama \cite{noza98}, which includes up to $O(\beta \theta_{e}^{2})$ terms.  The dashed curve is the present result that includes up to $O(\beta \theta_{e}^{3})$ terms.  The dot-dashed curve is the present result that includes up to $O(\beta \theta_{e}^{4})$ terms.  The solid curve is the present result of the direct numerical integration.}
\end{figure}

We have derived an analytic expression for the Sunyaev-Zeldovich effect in the Rayleigh--Jeans limit where $X \rightarrow 0$.  The result is
\begin{eqnarray}
\frac{\Delta n(X)}{n_{0}(X)} & \rightarrow & - 2 \tau \, \theta_{e} \, \left[ \, 1 - \frac{17}{10} \theta_{e} + \frac{123}{40} \theta_{e}^{2} - \frac{1989}{280} \theta_{e}^{3} + \frac{14403}{640} \theta_{e}^{4} \, \right]  \, \nonumber \\
 & & - 2 \tau \, \beta^{2} \left[ \, \frac{1}{3} \, - \, \frac{3}{10} \, \theta_{e} \,  + \frac{23}{20} \theta_{e}^{2} - \frac{2539}{560} \theta_{e}^{3} \,  \right]  \,   \nonumber  \\
 & &  + \, \, \tau \, \beta \, P_{1}({\rm cos} \theta_{\gamma}) \left[ \, 1 -  \frac{2}{5} \theta_{e} + \frac{13}{5} \theta_{e}^{2} \, - \frac{1689}{140} \theta_{e}^{3} + 
  \frac{7281}{140} \theta_{e}^{4} \, \right] \, \nonumber \\
 & & +  \, \, \tau \, \beta^{2} \, P_{2}({\rm cos} \theta_{\gamma}) \left[ \,  \frac{1}{15} \, - \, \frac{4}{5} \theta_{e} \, + \frac{34}{7} \theta_{e}^{2} - \frac{341}{14} \theta_{e}^{3}  \, \right] \, .
\end{eqnarray}

We have also carried out a direct numerical integration of the Boltzmann collision term on the right-hand-side of equation (14).  The method of the numerical integration is presented in Appendix.  Here we will show an example of our numerical calculations.  We show the results of the following case.  We have considered the case that $\theta_{e}$=0.02 and the cluster approaches the observer (cos$\theta_{\gamma}$=1) with peculiar velocity $v$=1000 kms$^{-1}$.  The numerical result for the kinematical Sunyaev-Zeldovich effect has been obtained by subtracting the numerically-obtained purely thermal effect ($\beta$=0) from the numerically-obtained total (thermal plus kinematical) effect.  Then we have compared this with our formula for the kinematical Sunyaev-Zeldovich effect (the sum of the second, third, and fourth lines of equation (20)) and also with Shimon and Rephaeli's formula for the kinematical Sunyaev-Zeldovich effect.

We define the following quantity
\begin{eqnarray}
\delta_{\rm present} & \equiv & \left| \frac{(\Delta I)_{\rm present \, \, analytic} - (\Delta I)_{\rm present \, \, numerical}}{(\Delta I)_{\rm present \, \, numerical}} \right| \, ,  \\
\delta_{\rm SR} & \equiv & \left| \frac{(\Delta I)_{\rm SR \, \, analytic} - (\Delta I)_{\rm present \, \, numerical}}{(\Delta I)_{\rm present \, \, numerical}} \right| \, . 
\end{eqnarray}
The results are shown in fig.\ 4.  It is readily seen that our analytic formula for the kinematical Sunyaev-Zeldovich effect agrees with the numerical results for small values of $X$ with accuracy better than $10^{-4}$.  On the other hand, Shimon and Rephaeli's formula shows about 2\% discrepancy with the numerical results for small values of $X$.  Therefore we have confirmed the excellent accuracy of our analytic formula for the kinematical Sunyaev-Zeldovich effect.  We conclude that Shimon and Rephaeli's formula for the kinematical Sunyaev-Zeldovich effect is in error.

\begin{figure}
\begin{center}
\includegraphics[angle=-90,width=0.8\textwidth]{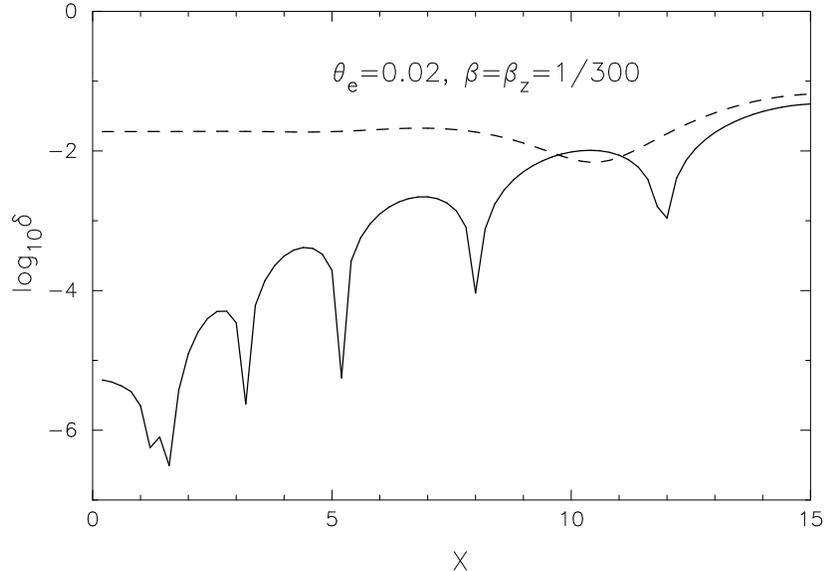}
\end{center}
\caption{The comparison of $\delta_{\rm present}$ (solid curve) and $\delta_{\rm SR}$ (dashed curve) as defined by equations (49) and (50), respectively.}
\end{figure}

A discussion on the discrepancy of Shimon and Rephaeli's results from ours is in order.  All of our calculations have been carried out consistently in the CMB frame.  The electron distribution function $f(E)$ in the CMB frame in equation (1) is equal to $f_{C}(E_{C})$, the electron distribution function in the CG frame (equation (6)).  The electron energy in the CG frame $E_{C}$ is expressed in terms of the electron energy and momentum in the CMB frame (equation (8)).  The integration is carried out consistently over the momentum space in the CMB frame (equation (1)).  However, Shimon and Rephaeli appear to have erroneously used the momentum space in the CG frame instead.  This is the likely cause of the discrepancy of Shimon and Rephaeli's results from ours.  Shimon and Rephaeli correctly reproduce the $C_{0}$ term in equation (20), and their discrepancy starts at the $\theta_{e} C_{1}$ term.  Since the contribution of the $\theta_{e} C_{1}$ term with respect to the contribution of the $C_{0}$ term for the case of $\theta_{e}$=0.02 is about 2\% for small values of $X$, this explains the results in fig.\ 4.

\section{Concluding remarks}

 We have extended the previous calculation of Nozawa, Itoh, \& Kohyama (1998) on the relativistic corrections to the kinematical Sunyaev-Zeldovich effect for clusters of galaxies to the order of $\beta \theta_{e}^{4}$ where $\theta_{e} \equiv k_{B}T_{e}/m_{e}c^{2}$.  We have also carried out a direct numerical integration of the Boltzmann collision term and numerically calculated the kinematical Sunyaev-Zeldovich effect, thereby confirming the excellent accuracy of our analytic formula for the kinematical Sunyaev-Zeldovich effect.  This precision formula should be very useful for the analysis of the observational data of the kinematical Sunyaev-Zeldovich effect for clusters of galaxies, which will be very likely to be obtained through the forthcoming CMB experiments.  We have found that Shimon and Rephaeli's formula for the kinematical Suyaev-Zeldovich effect contains about 2\% error for the case $\theta_{e}$=0.02, $\beta=\beta_{z}$=1/300 by comparing with the results obtained by the numerical integration of the Boltzmann collision term.

This is the first time that the accuracy of the analytic formula for the kinematical Sunyaev-Zeldovich effect has been quantitatively assessed by comparing with the results obtained by the numerical integration of the Boltzmann collision term.  It is gratifying to find that the excellent accuracy of the analytic formula for the kinematical Sunyaev-Zeldovich effect has been proved by comparing with the numerical results.

\acknowledgments

  We wish to thank our referee for many helpful suggestions.  One of the authors (N. I.) wishes to thank Lyman Page for a discussion on the upcoming ACT experiment, which has motivated him to produce a precision formula for the relativistic corrections to the kinematical Sunyaev-Zeldovich effect for clusters of galaxies.  He also wishes to thank Rashid Sunyaev and Jens Chluba for very inspiring discussions.  This work is financially supported in part by the Grants-in-Aid of the Japanese Ministry of Education, Culture, Sports, Science, and Technology under contracts \#15540293 and \#16540220.

\newpage

\appendix
\section*{}

  In this appendix we describe our method of the numerical integration of the Boltzmann collision term on the right-hand-side of equation (14).  Let us first start with the electron distribution function in the CG frame, i.e. equation (11).\begin{eqnarray}
f_{C}(E_{C}) & = & \left[ e^{\left\{(E_{C} - m)-(\mu_{C} - m) \right\}/k_{B}T_{e}} \, + \, 1 \right]^{-1}  \nonumber \\
& \approx & e^{-\left\{(E_{C}-m)-(\mu_{C} - m)\right\}/k_{B}T_{e}} \, ,
\end{eqnarray}
The electron number density $N_{e}$ is related to the electron distribution function as follows.
\begin{eqnarray}
N_e & = & 2\int_0^{\infty}\frac{d^{3}p_C}{(2\pi)^3}f_C\left(E_C\right) \nonumber \\
    & = & \frac{1}{\pi^2}e^{\left(\mu_C-m\right)/{k_BT_e}}
          \int_0^{\infty}dp_C\,\,p_C^2\,\,e^{-K_C/{k_BT_e}} \, ,
\end{eqnarray}
where $K_C \equiv E_C - m$.  This equation is further simplifed.
\begin{eqnarray}
N_e & = & e^{\left(\mu_C-m\right)/k_BT_e} \frac{m^3}{\pi^2} \theta_e \int_0^{\infty}du \left(1 + \theta_e u\right) \sqrt{\theta_e u \left(2 + \theta_e u\right)}\,\,e^{-u} \nonumber \, .
\end{eqnarray}
Therefore the chemical potential term in $f_{C}(E_{C})$ is expressed by $N_{e}$ as follows.
\begin{eqnarray}
e^{\left(\mu_C-m\right)/k_BT_e}
    & = &  N_e\frac{\pi^2}{m^3} \theta_e^{-\frac{3}{2}} \, g(\theta_e)  \, ,  \\
g(\theta_e) & \equiv & \left[ \int_0^{\infty}dx \left(1+ \theta_e x\right)\sqrt{x \left(2 + \theta_e x \right)}
                 \,\,e^{-x} \right]^{-1} \, .
\end{eqnarray}

  The equation (14) is rewritten with equation (2) as follows.
\begin{eqnarray}
\frac{\partial n(\omega)}{\partial t}
 &=& -2\left(\frac{e^2}{4\pi}\right)^2\int\frac{d^3p}{(2\pi)^3}\frac{d^3p'd^3k'\bar X}{2\omega\omega'EE'}
     \delta^4\left(p+k-p'-k'\right)f_C\left(E_C\right)\nonumber\\
 & & \times\left\{n(\omega)\left[1+n\left(\omega'\right)\right]-n\left(\omega'\right)\left[1+n(\omega)\right]
     e^{\Delta x\gamma\left(1-\vec\beta\cdot\hat k'\right)}
     e^{x\gamma\vec\beta\cdot\left(\hat k-\hat k'\right)}\right\} \nonumber \\
 & = &  -2\left(\frac{e^2}{4\pi}\right)^2\int\frac{d^3p}{(2\pi)^3}\frac{1}{E\omega}
     \int d\Omega_{k'}\int d\omega'\omega'\,\,\delta\left[\left(p+k-k'\right)^2-m^2\right]
     f_C\left(E_C\right){\bar X}\nonumber\\
 & & \times\left\{n(\omega)[1+n\left(\omega'\right)]-n\left(\omega'\right)[1+n(\omega)]
     e^{\Delta x\gamma\left(1-\vec\beta\cdot\hat k'\right)}
     e^{x\gamma\vec\beta\cdot\left(\hat k-\hat k'\right)}\right\}
\end{eqnarray}
Equation (A.5) is further simplified as follows.
\begin{eqnarray}
\frac{\partial n(\omega)}{\partial t}
 &=& -\left(\frac{e^2}{4\pi}\right)^2\int\frac{d^3p}{(2\pi)^3}\frac{1}{E}
     \int d\Omega_{k'}\frac{\omega'}{\omega}
     \frac{1}{E+\omega\left(1-\cos\theta_{\gamma'}\right)-p\cos\theta_p'}
     f_C\left(E_C\right){\bar X}\nonumber\\
 & & \times\left\{n(\omega)\left[1+n\left(\omega'\right)\right]-n\left(\omega'\right)\left[1+n(\omega)\right]
     e^{\Delta x\gamma\left(1-\vec\beta\cdot\hat k'\right)}
     e^{x\gamma\vec\beta\cdot\left(\hat k-\hat k'\right)}\right\}
\end{eqnarray}
and
\begin{eqnarray}
\omega'
 &=& \frac{\omega\left(E-p\cos\theta_p\right)}{E+\omega\left(1-\cos\theta_{\gamma'}\right)-p\cos\theta_p'}  \, ,
\end{eqnarray}
where $\theta_p$, $\theta_p'$ and $\theta_{\gamma'}$ are the angles between $\vec p$ and $\vec k$, $\vec p$ and $\vec k'$, $\vec k$ and $\vec k'$, respectively.  In equation (A.6) the chemical potential term of $f_{C}(E_{C})$ is replaced by equation (A.3).  

Finally we perform the five-dimensional integration of equation (A.6).  We use the Gaussian quadrature integration method, which is commonly used for the numerical integrations.  The method is very efficient as well as accurate.  We have checked the convergence of the integral carefully by varying the number of quadrature points.  The accuracy of the integral is $10^{-6}$ for the case of 70 quadrature points for each integral variable.  The numerical integration itself is straightforward, although it is rather time consuming.  It takes about 20 hours (CPU time) for a standard Alpha Server Workstation in order to complete the calculation of the spectral intensity $\Delta I$ for fixed ($\beta$, $\theta_{e}$) values and for $0 \leq X \leq 20$ with 100 points.

\end{document}